\documentclass[useAMS,usenatbib,usegraphicx]{mn2e}

\usepackage{color}

\title[Strong / Ultrastrong UV Fe {\sc II} Quasars]{Environments of
Strong / Ultrastrong, Ultraviolet Fe {\sc II} Emitting Quasars}

\author[R.G. Clowes et al.]
{Roger G.~Clowes,$^1$\thanks{E-mail: rgclowes@uclan.ac.uk}
Srinivasan Raghunathan,$^2$
Ilona K. S\"ochting,$^3$
\newauthor
Matthew J. Graham$^4$
and
Luis E. Campusano$^2$ \\
$^1$ Jeremiah Horrocks Institute, University of Central Lancashire,
Preston PR1 2HE \\
$^2$ Observatorio Astron\'omico Cerro Cal\'an, Departamento de Astronom\'{\i}a,
Universidad de Chile, Casilla 36-D, Santiago, Chile \\
$^3$ Astrophysics, Denys Wilkinson Building, Keble Road,
University of Oxford, Oxford OX1 3RH \\
$^4$ California Institute of Technology, 1200 East California Boulevard,
Pasadena, CA 91125, USA}

\begin{document}

\date{Accepted 2013 May 21. Received 2013 May 17; in original form
2013 February 13}

\pagerange{\pageref{firstpage}--\pageref{lastpage}} \pubyear{2011}

\maketitle

\label{firstpage}

\begin{abstract}
We have investigated the strength of ultraviolet Fe~{\sc II} emission from
quasars within the environments of Large Quasar Groups (LQGs) in comparison
with quasars elsewhere, for $1.1 \le \bar{z}_{LQG} \le 1.7$, using the DR7QSO
catalogue of the Sloan Digital Sky Survey. We use the Weymann et al.\ $W2400$
equivalent width, defined between the rest-frame continuum-windows 2240--2255
and 2665--2695~\AA, as the measure of the UV Fe~{\sc II} emission. We find a
significant shift of the $W2400$ distribution to higher values for quasars
within LQGs, predominantly for those LQGs with $1.1 \le \bar{z}_{LQG} \le
1.5$. There is a tentative indication that the shift to higher values
increases with the quasar $i$ magnitude. We find evidence that within LQGs
the ultrastrong emitters with $W2400 \ge 45$~\AA\ (more precisely,
ultrastrong-plus with $W2400 \ge 44$~\AA) have preferred nearest-neighbour
separations of $\sim$ 30--50~Mpc to the adjacent quasar of any $W2400$
strength. No such effect is seen for the ultrastrong emitters that are not in
LQGs. The possibilities for increasing the strength of the Fe~{\sc II}
emission appear to be iron abundance, Ly-$\alpha$ fluorescence, and
microturbulence, and probably all of these operate. The dense environment of
the LQGs may have led to an increased rate of star formation and an enhanced
abundance of iron in the nuclei of galaxies. Similarly the dense environment
may have led to more active blackholes and increased Ly-$\alpha$
fluorescence. The preferred nearest-neighbour separation for the stronger
emitters would appear to suggest a dynamical component, such as
microturbulence. In one particular LQG, the Huge-LQG (the largest structure
known in the early universe), six of the seven strongest emitters very
obviously form three pairings within the total of 73 members.
\end{abstract}


\section{Introduction}

From deep MMT / Hectospec spectroscopy to $g$ magnitude $\sim 21$,
\citet{Harris2012} and \citet{Harris2011} have serendipitously discovered a
relative strengthening of ultraviolet Fe~{\sc II} emission for quasars ($1.1
\le z \le 1.7$) in a 2 deg$^2$ ``pencil-beam'' field that intersects two
Large Quasar Groups (LQGs), U1.11 and U1.28 \citep{Clowes2012}, and a
``doubtful LQG'', U1.54. That work suggests that strong / ultrastrong UV Fe
{\sc II} emitters appear to be more strongly represented in dense quasar
environments, and that (Clowes \& Harris, unpublished visualisation) they
appear to clump with other quasars or with themselves. Such effects, if
confirmed, would be important for our understanding of the origin of the
problematic UV Fe~{\sc II} emission, the influence on it of large- and
small-scale environments, and of the evolution of cosmological structures. In
this paper, using SDSS (Sloan Digital Sky Survey) spectroscopy, we test
whether the above effects are observed when a large sample of LQGs is
considered. The pencil-beam study has deep spectroscopy but, of course,
narrow-angle coverage.  Here, with less-deep SDSS spectroscopy, we can
consider all of these LQGs in their entirety. We first discuss the
statistical properties of the UV Fe~{\sc II} for LQGs in general. We then
focus further on these three LQGs U1.11, U1.28, U1.54 together with the
``Huge-LQG'', U1.27, of \citet{Clowes2013}.

Large Quasar Groups are the largest structures seen in the early universe,
with sizes $\sim$ 70--500 Mpc and memberships of $\sim$ 5-70 quasars. See
\citet{Clowes2012} and \citet{Clowes2013}, and earlier references given
there, for more details on LQGs. The ``doubtful LQG'' U1.54 is a candidate
LQG that failed the test of significance described in \citet{Clowes2012}, but
which had been identified once before as a candidate LQG in an independent
survey \citep{Newman1998, Newman1999}. Sometimes, though, for simplicity we
shall refer to this doubtful LQG just as an LQG. Partly, we retain some
interest in U1.54 because it and U1.11 and U1.28 are all aligned along the
line of sight. The Huge-LQG \citep[U1.27,][]{Clowes2013} is the largest
structure currently known in the early universe. U1.28, also known as the
Clowes \& Campusano LQG, \citep[CCLQG,][]{Clowes1991, Clowes2012} was
previously the largest. (In this labelling of the LQGs, ``U'' refers to a
connected unit, and the appended number gives the mean redshift of its
members.)

Iron (Fe) in the optical and ultraviolet regions of quasar and AGN spectra is
very common, but its occurrence in the UV at the ``ultrastrong'' level is
rare. A good example of an ultrastrong, UV Fe~{\sc II} emitting quasar is
that of \citet*{Graham1996}, quasar 2226$-$3905. The total number of such
ultrastrong quasars known used to be very small, but it has now increased
because of the Sloan Digital Sky Survey (SDSS) --- for example,
\citet{Meusinger2012}, and our own work. The fractional rate of occurrence
remains small, however: we estimate that for redshifts in the range 1.0--1.8,
only $\sim$ 6.6 per cent of all quasars are ultrastrong emitters.
(Incidentally, we classify 88 per cent of the high-grade optical and/or UV
Fe~{\sc II} emitters with $i \le 19.1$ and $1.0 \le z \le 1.8$ from
\citet{Meusinger2012} as strong or ultrastrong UV emitters, and conversely we
find a substantially larger total.)

Following \citet{Weymann1991}, we use the rest-frame ``$W2400$'' equivalent
width, defined between two continuum-windows 2240--2255 and 2665--2695
\AA\ as the index of UV Fe~{\sc II} emission. Based on the median $W2400$
$\sim$ 30~\AA\ for all quasars (non-BAL and BAL) from Table~2 of
\citet{Weymann1991} we define \citep{Harris2012, Harris2011}, as an
illustrative guide, ``strong emitters'' as those with $30 \le W2400 <
45$~\AA\ and ``ultrastrong'' as those with $W2400 \ge 45$~\AA. Note, however
(see further discussion below) that the sample of $W2400$ values from
\citet{Weymann1991} seems to have substantially different properties from
ours.

The cause of ultrastrong, UV Fe~{\sc II} emission has not been successfully
attributed to any one mechanism, and probably the reality is that several
mechanisms contribute, particularly iron abundance, Ly-$\alpha$ fluorescence,
and microturbulence. \citet{Harris2011} gives a short but detailed review of
the current state of understanding. A very brief summary follows.

The Fe~{\sc II} is often attributed to the broad line region (BLR), but there
is evidence that it might instead arise from an intermediate line region
(ILR), between the outer BLR and the inner torus \citep[e.g.][]{Graham1996,
  Zhang2011}. In the early days of trying to understand the Fe~{\sc II}
emission, \citet*{Wills1985} and \citet{Collin-Souffrin1988} concluded that
either there was an unusually high abundance of iron or an important
mechanism was being overlooked. Abundance is important, but not likely to be
dominant \citep{Sigut2003, Baldwin2004}. \citet{Penston1987} proposed that
Ly$\alpha$ fluorescence might be the overlooked mechanism, with observational
and theoretical support following from \citet{Graham1996} and
\citet{Sigut1998} respectively. A further important mechanism is
microturbulence of $\sim 100$ km~s$^{-1}$ \citep{Ruff2012}, which increases
the spread in wavelength of Fe~{\sc II} absorption
\citep[e.g.][]{Bruhweiler2008}, and thus increases radiative pumping. Strong
and ultrastrong Fe~{\sc II} emission might be associated with some special
environmental circumstances influencing the quasars.

\citet{Harris2012} and \citet{Harris2011} find that there is a systematic
shift by $\sim$ 9~\AA\ of the $W2400$ equivalent widths to higher values for
the 2 deg$^2$ pencil-beam field that intersects the LQGs U1.11, U1.28 and
U1.54 compared with a combined set of 13 2-deg$^2$ control fields
elsewhere. There is then an unusually high rate of occurrence of strong and
ultrastrong Fe~{\sc II} emitters. These strong and ultrastrong emitters
appear to clump with other quasars or with themselves (Clowes \& Harris,
unpublished visualisation).

From the work of \citet{Harris2012} and \citet{Harris2011} it therefore
appears that strong and ultrastrong UV Fe~{\sc II} emitters might
preferentially occur where the density of quasars is high, as it is in
LQGs. In the case of the pencil-beam field there can be ambiguity about
membership of the LQGs by the MMT / Hectospec quasars since they are mostly
fainter than the $i \le 19.1$ quasars from which the LQGs were discovered.
The fainter quasars are considered provisionally to be new members primarily
if they fall within the convex hull \citep[of member spheres,][]{Clowes2012}
of the existing members. Even so, some are outside the LQGs, although they do
still clump --- with each other --- to form high-density regions. It is
possible that these quasars that are outside the LQGs would, with a fainter,
wide-angle survey prove to be members too, but with the data currently
available there is no way of knowing. It is also possible that the LQG
environment itself is not crucial, but that the LQGs are functioning as {\it
  providers} of the high (quasar) density environments that favour the Fe
{\sc II} emission. The alignment of these LQGs along the line of sight would
then presumably have made the environmental effect prominent and allowed its
discovery.

Of 14 strong / ultrastrong quasars in the pencil-beam field, that satisfy the
imposed signal-to-noise (s/n) criterion, four are known LQG members (all
necessarily from SDSS, but one has also been observed with MMT / Hectospec),
five are provisional new members (i.e.\ they fall within the convex hulls of
the LQGs), and five are non-members (all MMT / Hectospec). The one quasar
that fails the s/n criterion would be a non-member (from SDSS).

In this paper we first examine the statistical properties of $W2400$ emission
for all of the LQGs with $1.1 \le \bar{z}_{LQG} \le 1.7$ that we have
identified in the DR7QSO catalogue \citep{Schneider2010} for quasars with
$1.0 \le z \le 1.8$ and $i \le 19.1$. This allows us to investigate with a
much larger sample than that of \citet{Harris2012} and \citet{Harris2011} the
$W2400$ properties of LQG members compared with non-members --- that is, of
quasars that are known to be in dense (quasar) environments compared with
those that are not known to be in dense environments. We apply the condition
$1.1 \le \bar{z}_{LQG} \le 1.7$ to minimise any ``edge effects'' of
incomplete membership of the LQGs. We next investigate the distribution of
the nearest-neighbour separations of the strongest emitters in the LQGs
compared with the weaker emitters in the LQGs. Finally, we focus further on
the three LQGs U1.11, U1.28 and U1.54, together with the Huge-LQG, U1.27. We
consider here all of the members of these LQGs rather than only those
intersected by the pencil-beam. The SDSS spectroscopy for these LQGs is, of
course, not so deep as the MMT / Hectospec spectroscopy of the pencil-beam,
but the net outcome is nevertheless a useful increase in the size of the
strong / ultrastrong sample.

\section{The LQGs}

The LQGs have been discovered in the DR7QSO catalogue \citep{Schneider2010}
using the procedure described fully in \citet{Clowes2012} and summarised
briefly in \citet{Clowes2013}. Essentially, the procedure involves
application of a linkage algorithm followed by a test of statistical
significance, the CHMS significance \citep[where CHMS stands for convex hull
  of member spheres --- see][]{Clowes2012}. We restrict the DR7QSO quasars to
$i \le 19.1$ to ensure satisfactory spatial uniformity on the sky, since they
are then predominantly from the low-redshift strand of selection
\citep{Richards2006, Vanden-Berk2005}. We have selected LQG candidates from
quasars within the redshift range $1.0 \le z \le 1.8$, but for this work we
apply the condition $1.1 \le \bar{z}_{LQG} \le 1.7$ to minimise any edge
effects of incomplete membership. That is, the LQGs are restricted to $1.1
\le \bar{z}_{LQG} \le 1.7$ but the quasars within them are contained by $1.0
\le z \le 1.8$.

We consider here LQGs with CHMS-significance $\ge 2.8\sigma$ and number of
member quasars $\ge 10$. The choice of $2.8\sigma$ is such that contamination
by spurious LQG candidates should be negligible. This selection gives 134
LQGs with $1.0 \le z \le 1.8$, incorporating 3092 quasars in total. The
smallest membership is 11 and the largest 73 (U1.27, the Huge-LQG). The LQGs
are found from within a total of 27991 quasars. LQG members are thus only
$\sim$ 11 per cent of the total number of quasars. With the further condition
$1.1 \le \bar{z}_{LQG} \le 1.7$ we have 111 LQGs, incorporating 2629 quasars
in total. Again, the smallest membership is 11 and the largest 73. Details of
the LQGs will be published in a catalogue paper (Clowes, in preparation).

As mentioned above, the doubtful LQG, U1.54, is not formally a LQG since it
fails the test of CHMS-significance. It is therefore not one of the above 134
/ 111 LQGs that do pass the test. However, as mentioned above, we retain some
interest in U1.54 because it has been identified before as a candidate LQG
and because it, U1.11 and U1.28 are all aligned along the line of
sight. Also, its CHMS-significance is conservative, and, in the case of
curved morphology such as U1.54 has, there is therefore the possibility of a
candidate LQG being more interesting than is immediately apparent.

We shall focus further on the three LQGs U1.11, U1.28 and U1.54, together
with the Huge-LQG, U1.27. These are the LQGs that have so far been
investigated in most detail. Their properties are summarised in
Table~\ref{lqg_summary}.

\begin {table*}
\flushleft
\caption {Properties of the four LQGs U1.11, U1.28, U1.54 (the doubtful LQG)
and U1.27. The columns are: the name of the LQG; the number of member
quasars; the mean RA, Dec (2000); the mean redshift; the redshift range;
the characteristic size expressed as CHMS-volume$^{1/3}$ \citep[see][for
full details]{Clowes2012, Clowes2013}; and references. The total number
of member quasars is 166. They all have $i \le 19.1$, since this limit was
imposed on the DR7QSO catalogue \citep{Schneider2010} to ensure adequate
spatial uniformity of the data from which they were discovered. All were
discovered within the redshift range $1.0 \le z \le 1.8$. }
\small \renewcommand \arraystretch {0.8}
\newdimen\padwidth
\setbox0=\hbox{\rm0}
\padwidth=0.3\wd0
\catcode`|=\active
\def|{\kern\padwidth}
\newdimen\digitwidth
\setbox0=\hbox{\rm0}
\digitwidth=0.7\wd0
\catcode`!=\active
\def!{\kern\digitwidth}
\begin {tabular} {lllllll}
\\
LQG                     & Members & Mean RA, Dec (2000)       & Mean z & z range        & Volume$^{1/3}$ & References  \\
                        &         &                           &        &                & Mpc           &             \\
\\
U1.11                   & 38      & 10:46:13.9~~$+$03:27:10.4 & 1.11   & 1.0038--1.2007 & 380           & 2           \\
U1.28, CCLQG            & 34      & 10:49:10.3~~$+$05:17:09.0 & 1.28   & 1.1865--1.4232 & 350           & 1, 2        \\
U1.54, doubtful LQG     & 21      & 10:55:20.5~~$+$04:45:42.8 & 1.54   & 1.4765--1.6136 & 325           & 2           \\
U1.27, Huge-LQG         & 73      & 10:56:33.0~~$+$14:07:16.9 & 1.27   & 1.1742--1.3713 & 495           & 3           \\
\\
\end {tabular}
\\
References            \\
1 \citet{Clowes1991}  \\
2 \citet{Clowes2012}  \\
3 \citet{Clowes2013}  \\
\label{lqg_summary}
\end {table*}

\section{The W2400 measurements}

The member quasars of the LQGs all have $i \le 19.1$ and all are in the
redshift range $1.0 \le z \le 1.8$. Recall that to minimise edge effects we
are applying the condition $1.1 \le \bar{z}_{LQG} \le 1.7$. The rest-frame
equivalent width $W2400$, as described by \citet{Weymann1991}, has been
measured by software written for the purpose. It has been applied to all of
the DR7QSO quasars with $i \le 19.1$ and $1.0 \le z \le 1.8$, after smoothing
with a 5-pixel median filter.

The median filter is used for the following reason. The principal source of
error in the $W2400$ measurements seems likely to be the setting of the
continuum, given the quite narrow windows of the Weymann method --- 15 and 30
\AA\ in the rest frame for the two continuum-windows of 2240--2255 and
2665--2695~\AA. We have attempted an estimate of the measurement error by
comparing the $W2400$ values arising from the unsmoothed SDSS spectra with
those arising from smoothing (in the observed frame) with both a 5-pixel (as
used in the actual processing) and a 9-pixel median filter. The standard
deviation of the difference between them is $\sim$ 3.5 and 3.7
\AA\ respectively. The $W2400$ feature itself is so wide that the smoothing
of it by the median filter should have a relatively minor effect. We can
therefore adopt $\sim$ 3.5--3.7~\AA\ as an indicative error associated with
the $W2400$ measurements. We chose to apply the 5-pixel median filter
routinely for this purpose of setting the continuum levels more
reliably. (The 9-pixel median, however, would increase the effective width of
the lower continuum window in particular more than we would wish.)

The median filter also acts to reduce the residual [O~{\sc I}]
$\lambda$5577~\AA\ sky feature, for those spectra in which it is present. In
most cases, where present, its effect on $W2400$ is smaller than the
indicative errors. We can also assume that its occurrence in the LQG sample
is identical to that in the matched control sample (see below and the
following section).

Approximately 1 per cent of the measurements of $W2400$ by the software are
negative. Usually, this happens because the spectra are increasing strongly
to the blue, and are therefore concave, leading to negative $W2400$ values,
given the rigorous application of the $W2400$ definition. Occasionally,
absorption or artefacts in the spectra can also lead to negative values.

The 3092 LQG members ($1.0 \le z \le 1.8$) have been extracted from the
entire catalogue to form the LQG sample. The remainder --- 24899
non-LQG-members --- of the catalogue is then used as the control sample for
comparison.
The percentage occurrences of negative $W2400$ are very similar for both the
LQG sample and the control sample, and so negative $W2400$ values have simply
been removed, leading to final samples of 3063 and 24604 respectively. Also,
one of the 166 members of U1.11, U1.28, U1.54 and U1.27 has negative $W2400$.

For the final control sample, the mean and median $W2400$ values are 26.3 and
25.0~\AA\ respectively and the standard deviation is 12.2~\AA. The mean and
median here are thus lower than the median from \citet{Weymann1991} by $\sim$
5~\AA. The distribution of $W2400$ from \citet{Weymann1991} appears also to
be substantially different from ours, having no major symmetrical component
and having few low values. We suspect but cannot definitely establish that
the Weymann $W2400$ values are systematically too large. Perhaps the quasar
sample that they used is not representative of the quasar population. Another
possibility is that Weymann et al.\ used the {\sc IRAF splot} algorithm,
which was changed at about that time, leading to likely differences of $\sim$
15 per cent for wide, asymmetrical features (IRAF Newsletter no. 9,
1990). Conceivably, the differences could be larger still for a feature as
wide as the Fe~{\sc II} 2400~\AA\ emission.

Low-ionisation BAL quasars, showing Mg~{\sc II} BAL troughs, can lead to
$W2400$ values that are too large, since the continuum for the $W2400$
measurement is then underestimated. Such quasars can be removed using the
database of \citet{Shen2011}, but in practice the rate of occurrence is too
low to make anything other than a very slight difference to the analysis. The
\citet{Shen2011} database does not allow removal of all problematic Mg~{\sc
  II} absorption troughs, however, so, for the purpose of statistical
analysis, we simply assume that they can be neglected as their rate of
occurrence should be the same for both the LQG sample and the control sample.
(We also make the same assumption for the sample extracted for U1.11, U1.28,
U1.54 and U1.27.)

\section{Analysis of the W2400 distribution}

\begin{figure*}
\includegraphics[height=80mm]{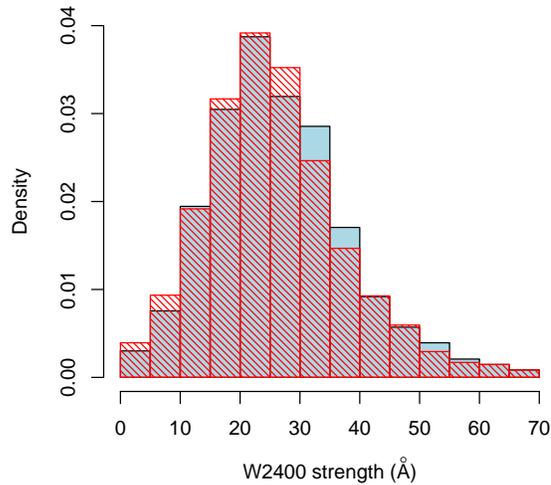}
\caption{The distribution of the rest-frame equivalent width, $W2400$, is
shown for the 111 LQGs with $1.1 \le \bar{z}_{LQG} \le 1.7$ as the solid
histogram (blue online, 2604 quasars). The distribution for the matched
control sample (see the text) is shown as the hatched histogram (red
online, 2604 quasars). Both are density histograms. Both are for $i \le
19.1$ and $1.0 \le z \le 1.8$, with the condition $1.1 \le \bar{z}_{LQG}
\le 1.7$ applied to the LQGs. The bin size is 5~\AA. As explained in the
text, negative $W2400$ values have been removed. The histograms have been
truncated at $W2400 = 70$~\AA\ for clarity.}
\label{w2400_all_lqgs}
\end{figure*}

The distribution of the rest-frame equivalent width, $W2400$, for the 2604
members of the 111 LQGs with $1.1 \le \bar{z}_{LQG} \le 1.7$ is shown in
Fig.~\ref{w2400_all_lqgs} as the solid histogram (blue online). The figure
shows in addition the distribution for a matched subset of the control sample
as the hatched histogram (red online). The matched subset, 
intended to negate possible dependences on magnitude and redshift, has
been created from the final control sample by finding, for each member of the
111 LQGs, the member of the control sample that is closest in $i$ magnitude
and in $10*z$, where $z$ is the redshift. The use of $10*z$ is so that we
give equal weight to an interval of 0.1 in $i$ and $0.01$ in $z$. Both
histograms are density histograms (meaning that relative
frequency is given by bin-height $\times$ bin-width). Both are for $i \le
19.1$ and $1.0 \le z \le 1.8$, with the condition $1.1 \le \bar{z}_{LQG} \le
1.7$ applied to the LQGs.  A one-sided Mann-Whitney test indicates that there
is a relative shift of the LQG-distribution to larger values at a level of
significance given by the p-value $= 0.0226$. The median shift is estimated
as 0.62~\AA. Note that the histograms of Fig.~\ref{w2400_all_lqgs} and
subsequent figures are for illustration, and the statistical analysis with
the Mann-Whitney test (or Kolmogorov-Smirnov test) does not depend on binned
data. If we restrict the $i$ magnitudes of the LQG quasars to $18.0 \le i \le
19.1$ then the p-value becomes $0.0064$ and the median shift becomes
0.84~\AA. The shift to higher $W2400$ thus appears to be a stronger effect at
fainter magnitudes.

\begin{figure*}
\includegraphics[height=80mm]{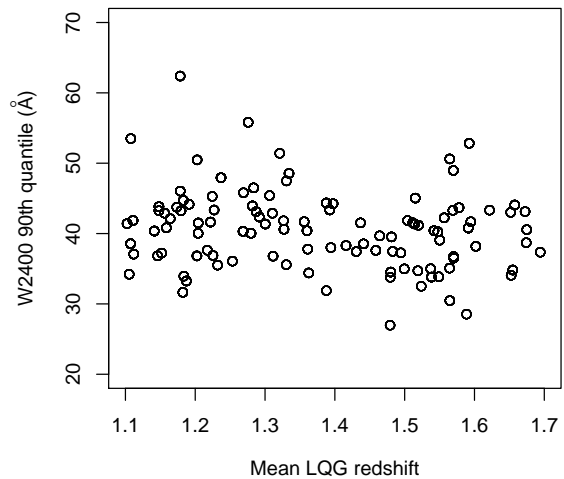}
\caption{A plot of the 90th quantile of the rest-frame equivalent
width, $W2400$, distribution against $\bar{z}_{LQG}$ for the 111 LQGs with
$1.1 \le \bar{z}_{LQG} \le 1.7$. Note the apparent change in properties
at $\bar{z}_{LQG} \sim 1.48$.}
\label{w2400q090_meanz}
\end{figure*}

The LQGs appear to show a change in the properties of their $W2400$
distributions at $\bar{z}_{LQG} \sim 1.48$. This change is illustrated in
Fig.~\ref{w2400q090_meanz} which plots, for each LQG with $1.1 \le
\bar{z}_{LQG} \le 1.7$, the value of the 90th quantile of the $W2400$
distribution against $\bar{z}_{LQG}$. The 90th quantile is used to
characterise the strong tail of the $W2400$ distribution for each LQG. There
appears to be a discontinuity in the typical value of the 90th quantile and
in the upper and lower envelopes at $\bar{z}_{LQG} \sim 1.48$.

\begin{figure*}
\includegraphics[height=80mm]{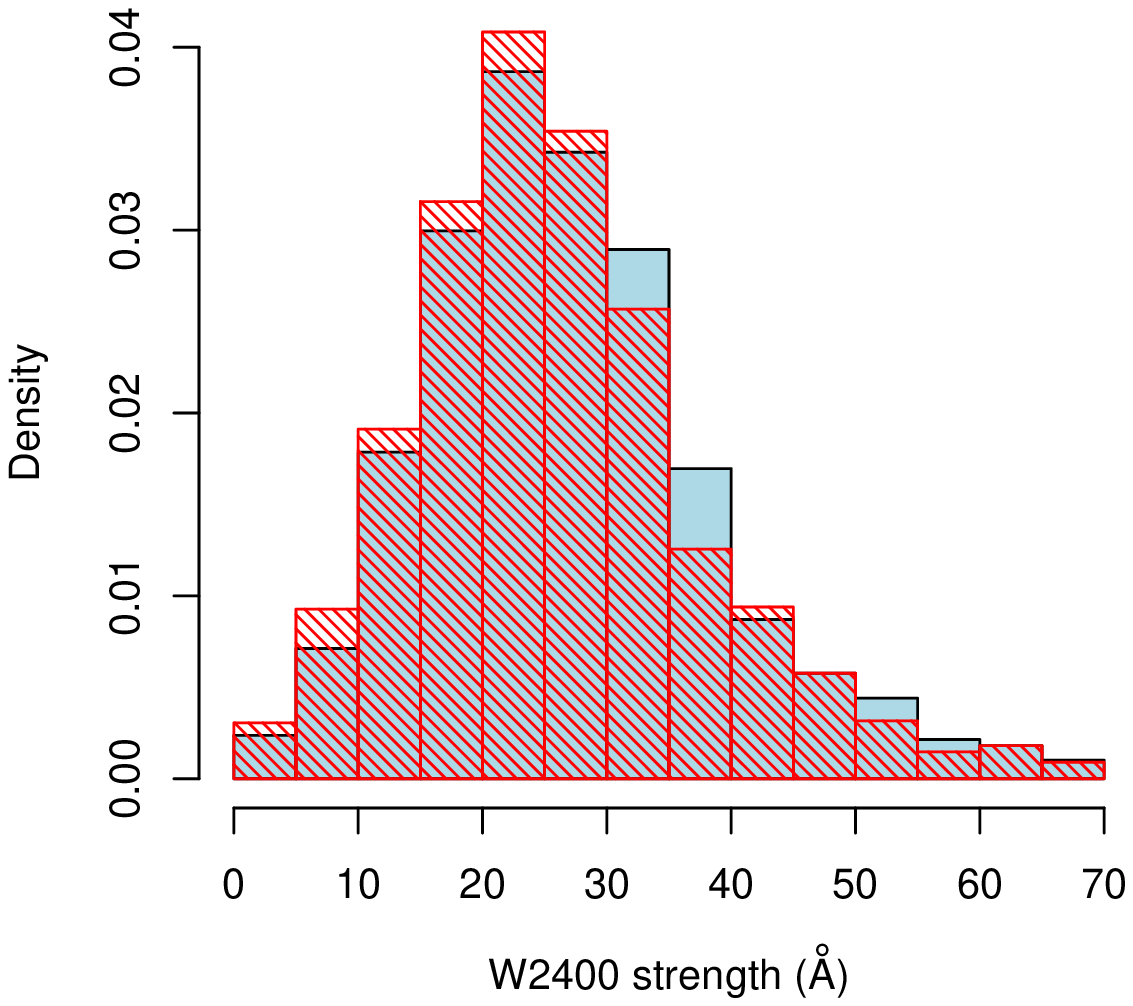}
\caption{The distribution of the rest-frame equivalent width, $W2400$, is
shown for the 75 LQGs with $1.1 \le \bar{z}_{LQG} \le 1.5$ as the solid
histogram (blue online, 1778 quasars). The distribution for the matched
control sample (see the text) is shown as the hatched histogram (red
online, 1778 quasars). Both are density histograms. Both are for $i \le
19.1$ and $1.0 \le z \le 1.8$, with the condition $1.1 \le \bar{z}_{LQG}
\le 1.5$ applied to the LQGs.  Other details are as for
Fig.~\ref{w2400_all_lqgs}.}
\label{w2400_all_lqgs_B}
\end{figure*}

Given this change at $\bar{z}_{LQG} \sim 1.48$, we apply instead the
condition $1.1 \le \bar{z}_{LQG} \le 1.5$. It then appears that the shift of
the LQG distribution to higher values of $W2400$ is strongly concentrated in
this redshift range. Fig.~\ref{w2400_all_lqgs_B} is similar to
Fig.~\ref{w2400_all_lqgs} but instead gives the distribution of the
rest-frame equivalent width, $W2400$, for the 1778 members of the 75 LQGs
with $1.1 \le \bar{z}_{LQG} \le 1.5$ as the solid histogram (blue
online). The hatched histogram (red online) shows the distribution for the
corresponding matched subset of the control sample. In this case the
one-sided Mann-Whitney test indicates a relative shift of the
LQG-distribution to larger values at p-value=$0.0042$. The median shift is
estimated as 0.97~\AA.  Note that this condition $1.1 \le \bar{z}_{LQG} \le
1.5$ leads to the member quasars having redshifts in the range $1.0006 \le z
\le 1.6093$. If we again restrict the $i$ magnitudes of the LQG quasars to
$18.0 \le i \le 19.1$ then the p-value becomes $0.00054$ and the median shift
becomes 1.31~\AA. Again, the shift to higher $W2400$ appears to be a stronger
effect at fainter magnitudes.

If, having considered the sub-range $1.1 \le \bar{z}_{LQG} \le 1.5$, we now
consider only the remaining $1.5 < \bar{z}_{LQG} \le 1.7$ then there is no
perceptible shift at all of the LQG-distribution to higher values of
$W2400$. Of course, the number of LQGs and members is then somewhat smaller
(36 LQGs, 826 members), but we cautiously conclude that the shift to higher
$W2400$ is indeed strongly concentrated in the range $1.1 \le \bar{z}_{LQG}
\le 1.5$.

In summary so far, we find a small but significant shift to higher values of
$W2400$ for LQGs with $1.1 \le \bar{z}_{LQG} \le 1.5$. This result contrasts
with that from \citet{Harris2012} and \citet{Harris2011} for a large shift
for quasars with $1.1 \le z \le 1.7$ in the pencil-beam field that intersects
the LQGs U1.11, U1.28 and U1.54. However, we do find some indication that the
shift that we find here increases to fainter magnitudes. The MMT / Hectospec
data used by \citet{Harris2012} and \citet{Harris2011} are for quasars that
are typically much fainter than those used here, so it may be that the
results can still be reconciled.

\begin{figure*}
\includegraphics[height=80mm]{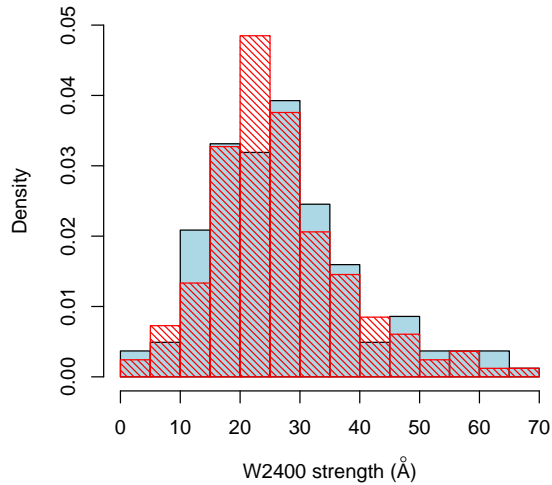}
\caption{The distribution of the rest-frame equivalent width, $W2400$, is
shown for the four LQGs U1.11, U1.28 (the CCLQG), U1.54 (the doubtful LQG)
and U1.27 (the Huge-LQG) as the solid histogram (blue online, 165 quasars).
The distribution for the matched control sample is shown as the hatched
histogram (red online, 165 quasars). Other details are as for
Fig.~\ref{w2400_all_lqgs}.}
\label{w2400_four_lqgs}
\end{figure*}

Fig.~\ref{w2400_four_lqgs} similarly shows the distribution of the rest-frame
equivalent width, $W2400$, for the four LQGs U1.11, U1.28 (the CCLQG) and
U1.54 (the doubtful LQG), and U1.27 (the Huge-LQG). Their $W2400$
distribution is shown as the solid histogram (blue online), and again the
distribution for the corresponding matched control sample is shown as the
hatched histogram (red online). The one-sided Mann-Whitney test indicates
that there is no significant shift to larger values. From the statistics for
the 75 LQGs with $1.1 \le \bar{z}_{LQG} \le 1.5$ we would, ignoring the
different redshift limits arising from the inclusion of U1.54, expect an
excess of only 1.6 ultrastrong and 5.8 strong emitters for these four LQGs
(166 quasars, 165 with positive $W2400$). Most probably the signal is lost
in the noise.

Note that, very unusually for the DR7QSO catalogue, nine of the 166 spectra
for U1.11, U1.27, U1.54 and U1.28 have exposure times of only 900~s. In fact,
only U1.11 and U1.28 are affected, with seven of the nine from U1.11 and two
from U1.28. Three of the nine are estimated to have s/n $< 4$ (two from U1.28
and one from U1.11). The software measurements of $W2400$ place two in the
ultrastrong category and one in the strong. Manual measurement suggests that,
despite the noise, the software measurements are acceptable.

The members of U1.11, U1.28, U1.54 and U1.27 that have been classified as
strong or ultrastrong emitters are listed in Table~\ref{SUS_table}. We can
briefly compare the classification here with those of \citet{Harris2011} for
the small area of the pencil-beam field. The superscripts in the first column
of Table~\ref{SUS_table} give the corresponding classification from
\citet{Harris2011} or, for SDSS J104932.22+050531.7, from Harris (private
communication). Note that SDSS J104938.35$+$052932.0 is classified as
ultrastrong here and weak in \citet{Harris2011}: the software measurement
here will be incorrect because of absorption occurring at the position of one
of the continuum windows. Two quasars classified as ultrastrong here but
strong by Harris are here on the boundary between the ultrastrong and strong
classifications. A quasar classified as strong by \citet{Harris2011} but weak
by the software measurements, SDSS J104840.34+055912.9 appears on manual
checking to be weak.

\begin {table*}
\flushleft
\caption {Strong ($30 \le W2400 < 45$~\AA) and ultrastrong ($W2400 \ge
45$~\AA) UV Fe~{\sc II} emitters in the four LQGs U1.11, U1.28, U1.54 and
U1.27 The columns are: category (whether strong or ultrastrong); quasar
SDSS name; redshift $z$; RA, Dec. (2000); the name of the LQG; $W2400$
equivalent width; $i$ magnitude.}
\small \renewcommand \arraystretch {0.8}
\newdimen\padwidth
\setbox0=\hbox{\rm0}
\padwidth=0.3\wd0
\catcode`|=\active
\def|{\kern\padwidth}
\newdimen\digitwidth
\setbox0=\hbox{\rm0}
\digitwidth=0.7\wd0
\catcode`!=\active
\def!{\kern\digitwidth}
\begin {tabular} {lllllll}
\\
Category         & Quasar SDSS name            & z      & RA, Dec (2000)              & LQG               & W2400  & $i$       \\
                 &                             &        &                             &                   & \AA    &           \\
\\
ultrastrong      & SDSS J104445.32$+$054348.8  & 1.1879 & 10:44:45.32~~$+$05:43:48.8  & U1.28 (CCLQG)     & 74.60  & 18.793    \\
ultrastrong$^w$  & SDSS J104938.35$+$052932.0  & 1.5169 & 10:49:38.35~~$+$05:29:32.0  & U1.54 (dou. LQG)  & 74.25  & 19.064    \\
ultrastrong      & SDSS J110412.00$+$044058.2  & 1.2554 & 11:04:12.00~~$+$04:40:58.2  & U1.28 (CCLQG)     & 67.30  & 18.851    \\
ultrastrong      & SDSS J110736.60$+$090114.7  & 1.2266 & 11:07:36.60~~$+$09:01:14.7  & U1.27 (Huge-LQG)  & 62.49  & 18.902    \\
ultrastrong      & SDSS J105527.67$+$002001.5  & 1.1448 & 10:55:27.67~~$+$00:20:01.5  & U1.11             & 62.17  & 18.782    \\
ultrastrong$^u$  & SDSS J104914.32$+$041428.6  & 1.6070 & 10:49:14.32~~$+$04:14:28.6  & U1.54 (dou. LQG)  & 61.17  & 18.871    \\
ultrastrong      & SDSS J104509.93$+$063559.0  & 1.1184 & 10:45:09.93~~$+$06:35:59.0  & U1.11             & 59.31  & 19.001    \\
ultrastrong      & SDSS J110810.87$+$014140.7  & 1.6136 & 11:08:10.87~~$+$01:41:40.7  & U1.54 (dou. LQG)  & 58.04  & 17.344    \\
ultrastrong      & SDSS J110504.46$+$084535.3  & 1.2371 & 11:05:04.46~~$+$08:45:35.3  & U1.27 (Huge-LQG)  & 55.20  & 19.005    \\
ultrastrong      & SDSS J104445.03$+$151901.6  & 1.2336 & 10:44:45.03~~$+$15:19:01.6  & U1.27 (Huge-LQG)  & 54.31  & 18.678    \\
ultrastrong      & SDSS J110121.37$+$054349.7  & 1.5252 & 11:01:21.37~~$+$05:43:49.7  & U1.54 (dou. LQG)  & 53.42  & 18.746    \\
ultrastrong      & SDSS J103744.89$+$051834.2  & 1.2280 & 10:37:44.89~~$+$05:18:34.2  & U1.28 (CCLQG)     & 52.54  & 18.958    \\
ultrastrong      & SDSS J104139.15$+$143530.2  & 1.2164 & 10:41:39.15~~$+$14:35:30.2  & U1.27 (Huge-LQG)  & 48.38  & 18.657    \\
ultrastrong      & SDSS J105224.08$+$204634.1  & 1.2032 & 10:52:24.08~~$+$20:46:34.1  & U1.27 (Huge-LQG)  & 47.74  & 18.593    \\
ultrastrong      & SDSS J103552.43$+$032537.2  & 1.0553 & 10:35:52.43~~$+$03:25:37.2  & U1.11             & 46.09  & 18.980    \\
ultrastrong$^s*$ & SDSS J104932.22$+$050531.7  & 1.1136 & 10:49:32.22~~$+$05:05:31.7  & U1.11             & 45.72  & 18.699    \\
ultrastrong      & SDSS J105119.60$+$142611.4  & 1.3093 & 10:51:19.60~~$+$14:26:11.4  & U1.27 (Huge-LQG)  & 45.57  & 19.002    \\
ultrastrong$^s$  & SDSS J105251.71$+$055733.7  & 1.5928 & 10:52:51.71~~$+$05:57:33.7  & U1.54 (dou. LQG)  & 45.11  & 18.440    \\
ultrastrong      & SDSS J104752.69$+$061828.9  & 1.3125 & 10:47:52.69~~$+$06:18:28.9  & U1.28 (CCLQG)     & 45.07  & 18.954    \\
strong$^+$       & SDSS J105144.88$+$125828.9  & 1.3153 & 10:51:44.88~~$+$12:58:28.9  & U1.27 (Huge-LQG)  & 44.48  & 19.021    \\
strong           & SDSS J105534.66$+$033028.8  & 1.2495 & 10:55:34.66~~$+$03:30:28.8  & U1.28 (CCLQG)     & 41.25  & 18.195    \\
strong           & SDSS J110016.88$+$193624.7  & 1.2399 & 11:00:16.88~~$+$19:36:24.7  & U1.27 (Huge-LQG)  & 40.41  & 18.605    \\
strong           & SDSS J103626.33$+$045436.4  & 1.0477 & 10:36:26.33~~$+$04:54:36.4  & U1.11             & 40.20  & 18.404    \\
strong           & SDSS J105611.27$+$170827.5  & 1.3316 & 10:56:11.27~~$+$17:08:27.5  & U1.27 (Huge-LQG)  & 39.87  & 17.698    \\
strong           & SDSS J105537.63$+$040520.0  & 1.2619 & 10:55:37.63~~$+$04:05:20.0  & U1.28 (CCLQG)     & 38.98  & 18.651    \\
strong           & SDSS J105821.28$+$053448.9  & 1.2540 & 10:58:21.28~~$+$05:34:48.9  & U1.28 (CCLQG)     & 38.87  & 18.134    \\
strong           & SDSS J104012.14$+$043904.6  & 1.1195 & 10:40:12.14~~$+$04:39:04.6  & U1.11             & 38.57  & 18.578    \\
strong           & SDSS J105525.68$+$113703.0  & 1.2893 & 10:55:25.68~~$+$11:37:03.0  & U1.27 (Huge-LQG)  & 38.54  & 18.264    \\
strong           & SDSS J110217.19$+$083921.1  & 1.2355 & 11:02:17.19~~$+$08:39:21.1  & U1.27 (Huge-LQG)  & 37.75  & 18.800    \\
strong           & SDSS J105141.89$+$045831.8  & 1.6080 & 10:51:41.89~~$+$04:58:31.8  & U1.54 (dou. LQG)  & 37.70  & 18.906    \\
strong           & SDSS J111823.21$+$090504.9  & 1.1923 & 11:18:23.21~~$+$09:05:04.9  & U1.27 (Huge-LQG)  & 37.65  & 18.940    \\
strong           & SDSS J105132.22$+$145615.1  & 1.3607 & 10:51:32.22~~$+$14:56:15.1  & U1.27 (Huge-LQG)  & 36.16  & 18.239    \\
strong           & SDSS J104116.79$+$035511.4  & 1.2444 & 10:41:16.79~~$+$03:55:11.4  & U1.28 (CCLQG)     & 35.74  & 18.531    \\
strong           & SDSS J103748.36$+$040242.1  & 1.0869 & 10:37:48.36~~$+$04:02:42.1  & U1.11             & 35.56  & 17.857    \\
strong           & SDSS J105352.72$+$050043.9  & 1.1320 & 10:53:52.72~~$+$05:00:43.9  & U1.11             & 35.39  & 18.865    \\
strong           & SDSS J105719.23$+$045548.2  & 1.3355 & 10:57:19.23~~$+$04:55:48.2  & U1.28 (CCLQG)     & 35.36  & 18.429    \\
strong           & SDSS J104656.71$+$054150.3  & 1.2284 & 10:46:56.71~~$+$05:41:50.3  & U1.28 (CCLQG)     & 34.66  & 17.594    \\
strong           & SDSS J104843.05$+$064456.8  & 1.3523 & 10:48:43.05~~$+$06:44:56.8  & U1.28 (CCLQG)     & 34.55  & 18.721    \\
strong           & SDSS J104410.13$+$072305.6  & 1.1514 & 10:44:10.13~~$+$07:23:05.6  & U1.11             & 34.51  & 18.189    \\
strong           & SDSS J104430.92$+$160245.0  & 1.2294 & 10:44:30.92~~$+$16:02:45.0  & U1.27 (Huge-LQG)  & 34.29  & 17.754    \\
strong           & SDSS J105637.49$+$150047.5  & 1.3713 & 10:56:37.49~~$+$15:00:47.5  & U1.27 (Huge-LQG)  & 33.58  & 19.041    \\
strong           & SDSS J105833.86$+$055440.2  & 1.3222 & 10:58:33.86~~$+$05:54:40.2  & U1.28 (CCLQG)     & 33.40  & 18.758    \\
strong           & SDSS J105621.90$+$143401.0  & 1.2333 & 10:56:21.90~~$+$14:34:01.0  & U1.27 (Huge-LQG)  & 33.37  & 19.052    \\
strong           & SDSS J105832.01$+$170456.0  & 1.2813 & 10:58:32.01~~$+$17:04:56.0  & U1.27 (Huge-LQG)  & 33.36  & 18.299    \\
strong           & SDSS J105017.31$+$012450.9  & 1.2007 & 10:50:17.31~~$+$01:24:50.9  & U1.11             & 33.07  & 18.800    \\
strong           & SDSS J104425.80$+$060925.6  & 1.2523 & 10:44:25.80~~$+$06:09:25.6  & U1.28 (CCLQG)     & 32.40  & 18.652    \\
strong           & SDSS J104114.06$+$034312.0  & 1.2633 & 10:41:14.06~~$+$03:43:12.0  & U1.28 (CCLQG)     & 32.20  & 18.588    \\
strong           & SDSS J104309.70$+$075317.8  & 1.1823 & 10:43:09.70~~$+$07:53:17.8  & U1.11             & 32.19  & 18.872    \\
strong           & SDSS J103639.63$+$022553.5  & 1.0525 & 10:36:39.63~~$+$02:25:53.5  & U1.11             & 32.17  & 18.817    \\
strong           & SDSS J105442.71$+$104320.6  & 1.3348 & 10:54:42.71~~$+$10:43:20.6  & U1.27 (Huge-LQG)  & 31.82  & 18.844    \\
strong           & SDSS J105022.81$+$064621.8  & 1.2900 & 10:50:22.81~~$+$06:46:21.8  & U1.28 (CCLQG)     & 31.07  & 18.362    \\
strong           & SDSS J105512.23$+$061243.9  & 1.3018 & 10:55:12.23~~$+$06:12:43.9  & U1.28 (CCLQG)     & 31.00  & 18.413    \\
strong           & SDSS J105525.18$+$191756.3  & 1.2005 & 10:55:25.18~~$+$19:17:56.3  & U1.27 (Huge-LQG)  & 30.81  & 18.833    \\
strong           & SDSS J104954.70$+$160042.3  & 1.3373 & 10:49:54.70~~$+$16:00:42.3  & U1.27 (Huge-LQG)  & 30.78  & 18.748    \\
strong           & SDSS J105541.83$+$111754.2  & 1.3298 & 10:55:41.83~~$+$11:17:54.2  & U1.27 (Huge-LQG)  & 30.57  & 18.996    \\
strong           & SDSS J105245.80$+$134057.4  & 1.3544 & 10:52:45.80~~$+$13:40:57.4  & U1.27 (Huge-LQG)  & 30.28  & 18.211    \\
\\
\end {tabular}
\\
$^u$   Classified as ultrastrong in \citet{Harris2011}.                            \\
$^s$   Classified as strong in \citet{Harris2011}.                                 \\
$^s*$  Classified as strong (Harris, private communication).                       \\
$^w$   Classified as weak ($W2400 < 30$~\AA) in \citet{Harris2011}.                \\
$^+$   This and entries above are classified also as ultrastrong-plus
       ($W2400 \ge 44$~\AA) --- see the text.                                      \\
\label{SUS_table}
\end {table*}

\section{Environments of the ultrastrong UV Fe~{\sc II} emitters}

In this section we discuss the environments of the ultrastrong ($W2400 \ge
45$~\AA) UV Fe~{\sc II} emitters compared with the weak ($W2400 < 30$~\AA)
within the 75 LQGs having $1.1 \le \bar{z}_{LQG} \le 1.5$. We also briefly
discuss the particular LQGs U1.11, U1.28, U1.54 and U1.27. If the shift to
higher $W2400$ within the LQGs arises from an environmental effect then we
might anticipate there will be a density or clustering effect that appears
most strongly for the ultrastrong emitters compared with the weak. There is
an arbitary element to this approach, of course, because the definitions of
ultrastrong, strong and weak emitters involve arbitrary boundaries.

Clowes \& Harris (unpublished visualisation) found that the strong and
ultrastrong emitters tended to clump with other quasars or with themselves,
both within and outside the convex hulls of the three LQGs U1.11, U1.28 and
U1.54. (Recall that the deep MMT / Hectospec spectroscopy in the pencil-beam
included quasars fainter than the $i \le 19.1$ limit of the quasars used for
the discovery of the LQGs.) Although this visualisation involved no
quantification of the clumping, it has guided our thinking that there could
be a preferred nearest-neighbour scale for the strongest emitters.

For these reasons, we have investigated the distribution of nearest-neighbour
separations for the ultrastrong emitters within the LQGs compared with the
distribution for the weak emitters within the LQGs. The nearest neighbours
have been determined without regard for the strength of their UV Fe~{\sc II}
emission. We consider only the LQGs with $1.1 \le \bar{z}_{LQG} \le 1.5$,
given the earlier result that the shift to higher values of $W2400$ is
strongly concentrated in this range.

\begin{figure*}
\includegraphics[height=80mm]{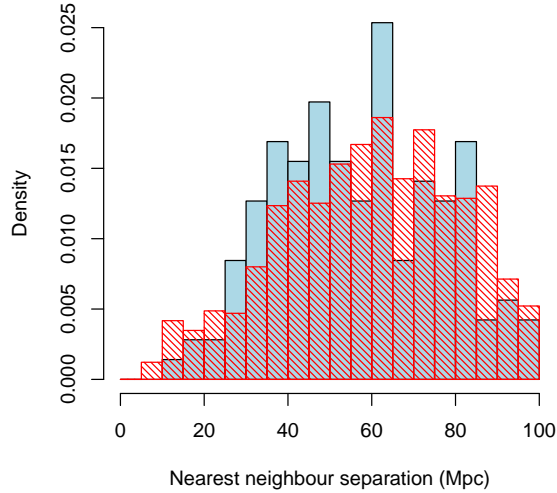}
\caption{The distribution of the nearest-neighbour separations (present
epoch) for the members of the 75 LQGs with $1.1 \le \bar{z}_{LQG} \le 1.5$
and for $W2400 \ge 45$~\AA\ (i.e.\ ultrastrong) is shown as the solid
histogram (blue online, 143 quasars), with mean $57.46 \pm
1.70$~Mpc. The distribution for the members of the LQGs with $W2400 <
30$~\AA\ is shown as the hatched histogram (red online, 1152 quasars),
with mean $59.65 \pm 0.63$~Mpc. Both are density
histograms. Both are for $i \le 19.1$ and $1.0 \le z \le 1.8$. The bin size
is 5~Mpc. Note the apparent preference of the $W2400 \ge 45$~\AA\ quasars
for separations in the range 25--50~Mpc. Note that the LQG-finding
algorithm restricts separations to $\le 100$~Mpc.}
\label{w2400nnsep_all_lqgs}
\end{figure*}

The distribution of the nearest-neighbour separations (present epoch) for the
members of the 75 LQGs with $1.1 \le \bar{z}_{LQG} \le 1.5$ and for $W2400
\ge 45$~\AA\ (i.e.\ ultrastrong) is shown in Fig.~\ref{w2400nnsep_all_lqgs}
as the solid histogram (blue online), with mean $57.46 \pm
1.70$~Mpc. The figure shows also the corresponding distribution for $W2400
< 30$~\AA\ (i.e.\ weak) as the hatched histogram (red online), 
with mean $59.65 \pm 0.63$~Mpc. Both are density histograms. Both are for
$i \le 19.1$ and $1.0 \le z \le 1.8$. Although drawn from the same LQGs, the
two histograms appear different, with that for $W2400 \ge 45$~\AA\ indicating
preferred values of the nearest-neighbour separation predominantly in the
range $\sim$ 25--50~Mpc (present epoch) --- note the consecutive
bins where the histogram for $W2400 \ge 45$~\AA\ exceeds the histogram for
$W2400 < 30$~\AA. A Mann-Whitney test is inappropriate here because the
LQG-finding algorithm restricts separations to $\le$ 100~Mpc, but a one-sided
Kolmogorov-Smirnov test indicates that the CDF (cumulative distribution
function) of the $W2400 \ge 45$~\AA\ distribution is greater than the CDF of
the $W2400 < 30$~\AA\ distribution, with p-value $=0.0255$, which is
marginally significant.

\begin{figure*}
\includegraphics[height=80mm]{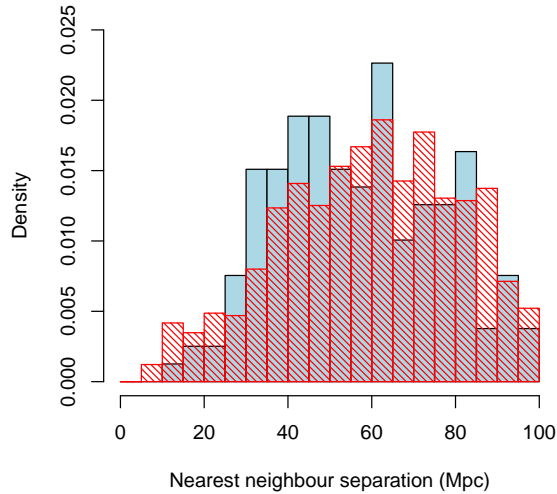}
\caption{The distribution of the nearest-neighbour separations (present
epoch) for the members of the 75 LQGs with $1.1 \le \bar{z}_{LQG} \le 1.5$
and for $W2400 \ge 44$~\AA\ (post-hoc adjustment) is shown as the solid
histogram (blue online, 160 quasars), with mean $57.40 \pm
1.60$~Mpc. The distribution for the members of the LQGs with $W2400 <
30$~\AA\ is shown as the hatched histogram (red online, 1152 quasars),
with mean $59.65 \pm 0.63$~Mpc. Both are density
histograms. Both are for $i \le 19.1$ and $1.0 \le z \le 1.8$. The bin size
is 5~Mpc. Note the apparent preference of the $W2400 \ge 44$~\AA\ quasars
for separations in the range 30--50~Mpc. Note that the LQG-finding
algorithm restricts separations to $\le 100$~Mpc.}
\label{w2400nnsep_all_lqgs_B}
\end{figure*}

Given the arbitrary boundaries for the definitions of ultrastrong, strong and
weak emitters we note that a post-hoc adjustment from $W2400 \ge 45$~\AA\ to
$W2400 \ge 44$~\AA\ gives the histograms shown in
Fig.~\ref{w2400nnsep_all_lqgs_B}, with means $57.40 \pm 1.60$,
$59.65 \pm 0.63$~Mpc, and the Kolmogorov-Smirnov test gives a p-value
$=0.0197$. The preferred nearest-neighbour separation appears then to be
predominantly in the range $\sim$ 30--50~Mpc.

Although we find evidence for a preferred nearest-neighbour scale for the
ultrastrong-emitting quasars in LQGs we find no evidence for a 
preferred scale in the quasars that are not members of LQGs 
(mean nearest-neighbour separation $\sim$ 78~Mpc), across a comparable range
of redshifts. Thus the preferred nearest-neighbour scale of $\sim$ 30--50~Mpc
for the $W2400 \ge 45$~\AA\ (more precisely, $W2400 \ge 44$~\AA) emitters
seems to be peculiar to the LQG environment. Presumably, it is related to the
shift to higher $W2400$ values within the LQGs.

\begin{figure*}
\includegraphics[height=100mm]{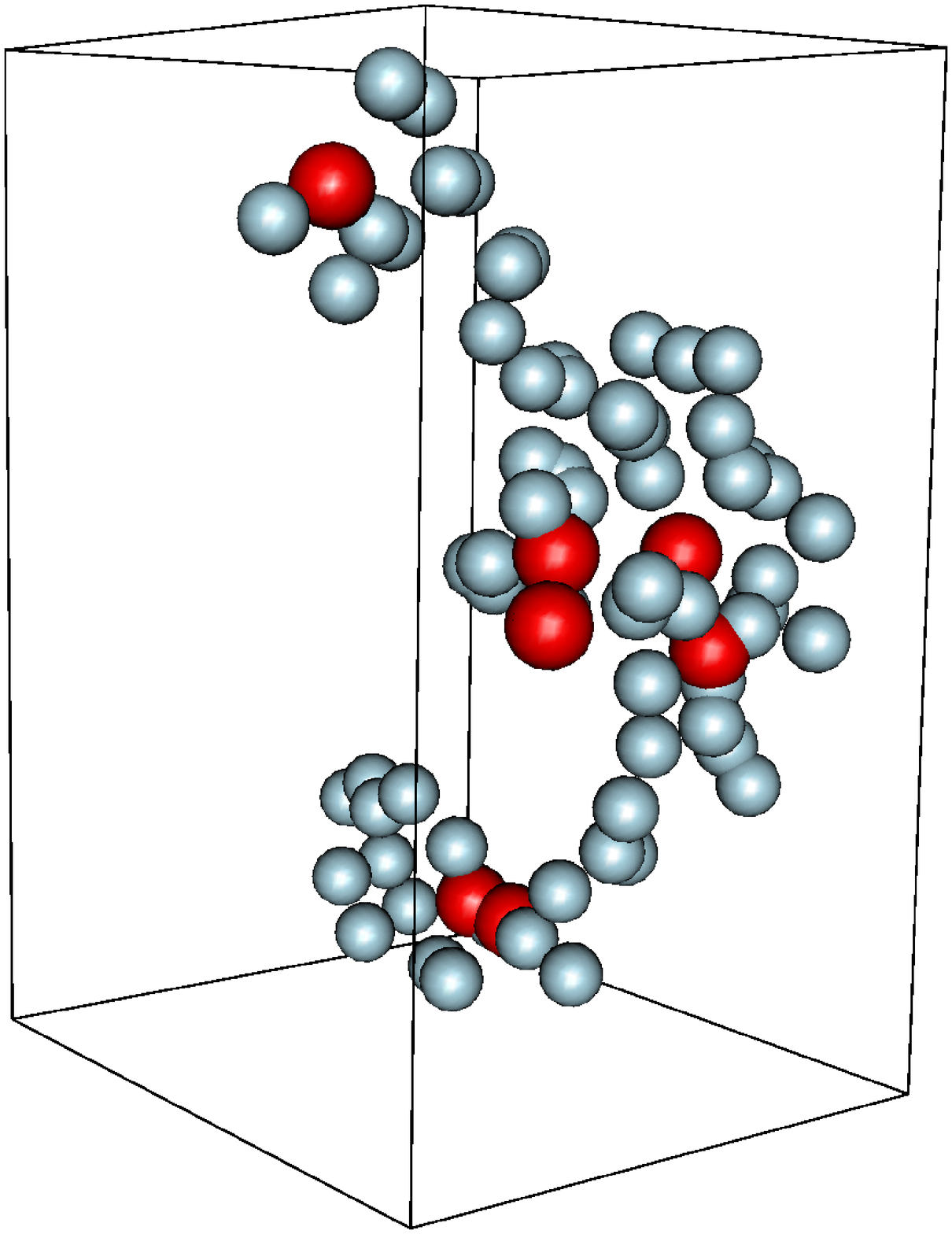}
\caption{Visualisation of the location of the seven ultrastrong-plus emitters
($W2400 \ge 44$~\AA) within U1.27, the Huge-LQG. The LQG has 73
members. Sixty-six of the members are represented by spheres of radius that
corresponds to 33~Mpc (present epoch, blue online). The seven
ultrastrong-plus quasars are represented by slightly larger spheres of radius
that corresponds to 40~Mpc (red online). The long dimension of the
surrounding box corresponds to approximately 1000~Mpc.}
\label{visU1.27_vsmall}
\end{figure*}

We have similarly used the visualisation software to look at the environments
of the strongest emitters from the entire memberships of the same three LQGs
from \citet{Harris2012} and \citet{Harris2011}, U1.11, U1.28 and U1.54,
together with U1.27. Given the above result on the nearest-neighbour
separations we concentrate on the quasars with $W2400 \ge 44$~\AA\ rather
than simply $W2400 \ge 45$~\AA. To simplify the discussion we introduce an
ultrastrong-plus category as those with $W2400 \ge 44$~\AA\ (while still
retaining ``ultrastrong'' as those with $W2400 \ge 45$~\AA). The findings are
similar to those previously, but not always so clear, perhaps because of the
brighter limiting magnitude. For example, with U1.54, the doubtful LQG, only
two of the five ultrastrong-plus emitters appear to be closely associated
with other quasars, and only one of these is in a dense region. In contrast,
with U1.11, three of four ultrastrong-plus emitters do seem to be associated
with other quasars and denser regions. With U1.28, two of four
ultrastrong-plus emitters appear to be close to other quasars, but with
neither being in particularly dense regions. With U1.27 (the Huge-LQG), six
of the seven ultrastrong-plus emitters form three pairs (separations 52, 81,
105~Mpc), within the total membership of 73. One pairing is actually part of
a close triplet with another quasar in a generally dense region, and another
pairing is part of a looser triplet in a less dense region. The third pairing
has one quasar in a dense region and the other detached from it. The seventh
ultrastrong-plus quasar is close to another quasar in a less dense region. A
visualisation of the location of these seven ultrastrong-plus emitters within
U1.27 is shown in Fig.~\ref{visU1.27_vsmall}.

\section{Discussion and conclusions}

We have compared the distribution of the $W2400$ equivalent width of UV Fe
{\sc II} emitting quasars in the dense environments of 111 LQGs having $1.1
\le \bar{z}_{LQG} \le 1.7$ with the distribution for quasars that are not in
LQGs. We find a marginally significant shift (p-value $= 0.0226$, shift
$=0.62$~\AA) of the $W2400$ distribution to higher values for quasars within
LQGs. The shift appears to be a stronger effect at fainter magnitudes
$18.0 \le i \le 19.1$ (p-value $= 0.0064$, shift $=0.84$~\AA).

However, the shift to higher $W2400$ appears to be strongly concentrated in
the 75 LQGs having $1.1 \le \bar{z}_{LQG} \le 1.5$ (p-value $= 0.0042$, shift
$=0.97$~\AA), and again it appears to be a stronger effect at fainter
magnitudes $18.0 \le i \le 19.1$ (p-value $= 0.00054$, shift $=1.31$~\AA).

We have investigated the distribution of nearest-neighbour separations for
the ultrastrong emitters within the 75 LQGs having $1.1 \le \bar{z}_{LQG} \le
1.5$ compared with the distribution for the weak emitters. The nearest
neighbours were determined without regard for the strength of their UV Fe
{\sc II} emission.  We found a marginally significant result (p-value
$=0.0255$) that the CDF of the ultrastrong ($W2400 \ge 45$~\AA) distribution
is greater than the CDF of the weak ($W2400 < 30$~\AA) distribution. There
appears to be a preferred nearest-neighbour separation predominantly in the
range $\sim$ 25--50~Mpc.  We found that a post-hoc adjustment from
ultrastrong $W2400 \ge 45$~\AA\ to ``ultrastrong-plus'' $W2400 \ge
44$~\AA\ gives improved significance (p-value $=0.0197$), and a preferred
nearest-neighbour separation that then appears to be predominantly in the
range $\sim$ 30--50~Mpc.

Our finding of a shift towards higher values of $W2400$ for quasars within
LQGs is compatible with the result from \citet{Harris2012} and
\citet{Harris2011} of a shift for a deep pencil-beam field that intersects
the three LQGs U1.11, U1.28 and U1.54 (the doubtful LQG). Our shift seems
smaller, but we do find an indication that the shift will increase to fainter
magnitudes. The MMT / Hectospec data used by \citet{Harris2012} and
\citet{Harris2011} are for quasars that are typically much fainter than those
used here, so it may be that the size of the shifts can still be
reconciled. However, there is an inevitable difficulty in constructing a
matching control sample for the faint MMT / Hectospec sample, so possibly the
attempted allowance for the differences has not been wholly successful and
the size of the shift there then appears larger than it should be. We find
that the shift is strongly concentrated in the range $1.1 \le \bar{z}_{LQG}
\le 1.5$. U1.54 is excluded from our data by this condition of course, but
fundamentally it is excluded because it fails our CHMS-significance
criterion.

Clowes \& Harris (unpublished visualisation) noted, for the pencil-beam field
within U1.11, U1.28 and U1.54, a tendency for strong / ultrastrong emitters
to clump with other quasars or with themselves. With the data for the 75 LQGs
having $1.1 \le \bar{z}_{LQG} \le 1.5$ in this paper we find evidence that
ultrastrong-plus emitters --- those with $W2400 \ge 44$~\AA\ --- have a
preferred nearest-neighbour scale of $\sim$ 30--50~Mpc. This result supports,
and makes quantitative, the result from the visualisation. However, no such
preferred scale is seen for ultrastrong-plus emitters that are not in LQGs.

Our visualisation of the ultrastrong-plus emitters across U1.11, U1.28 and
U1.54 in their entireties, together with U1.27, is generally consistent with
the deeper pencil-beam visualisation but not so clear. A striking result,
however, is from U1.27, the Huge-LQG, for which six of the seven
ultrastrong-plus emitters, from amongst the total of 73 LQG members, form three
pairings.

We have thus found two effects of the dense LQG environment on the
ultraviolet Fe~{\sc II} emission of the member quasars. Firstly, there is a
general shift to higher $W2400$, compared with non-members for $1.1 \le
\bar{z}_{LQG} \le 1.5$. The shift appears to be stronger for fainter
magnitudes. This redshift dependence, $\bar{z}_{LQG} \le 1.5$, suggests an
evolutionary effect. Secondly, we find evidence for a preferred
nearest-neighbour separation of $\sim$ 30--50~Mpc for the ultrastrong ($W2400
\ge 45$~\AA) or, more precisely, ultrastrong-plus ($W2400 \ge 44$~\AA)
emitters compared with the weak ($W2400 < 30$) emitters within these
LQGs. This preferred separation suggests a clustering or dynamical effect.
Of course, there may be further subtleties present in the dependences on
redshift, magnitude and density than we have seen, but a still larger sample
with which to disentangle them is not currently feasible. It may
be, however, that a different approach on the existing data, such as
fitting templates to the Fe~{\sc II}, could restore some information that
might presently be lost to uncertainties in measuring $W2400$.

The possibilities for increasing the strength of the Fe~{\sc II} emission
appear to be iron abundance, Ly-$\alpha$ fluorescence, and microturbulence.
Probably all of these operate. The dense environment of the LQGs and an
increased rate of interactions and mergers between galaxies may have led to
an increased rate of star formation and an enhanced abundance of iron in the
nuclei of galaxies. Similarly the dense environment of the LQGs may have led
to more active blackholes and increased Ly-$\alpha$ emission. The preferred
nearest-neighbour separation for the stronger emitters would appear to
suggest a dynamical component, such as microturbulence.

\section{Acknowledgments}

The anonymous referee is thanked for helpful comments and suggestions.

LEC received partial support from the Center of Excellence in Astrophysics
and Associated Technologies (PFB 06), and from a CONICYT Anillo project (ACT
1122).

SR is in receipt of a CONICYT PhD studentship.


This research has used the SDSS DR7QSO catalogue \citep{Schneider2010}.

Funding for the SDSS and SDSS-II has been provided by the Alfred P. Sloan
Foundation, the Participating Institutions, the National Science
Foundation, the U.S. Department of Energy, the National Aeronautics and
Space Administration, the Japanese Monbukagakusho, the Max Planck
Society, and the Higher Education Funding Council for England. The SDSS
Web Site is http://www.sdss.org/.

The SDSS is managed by the Astrophysical Research Consortium for the
Participating Institutions. The Participating Institutions are the American
Museum of Natural History, Astrophysical Institute Potsdam, University of
Basel, University of Cambridge, Case Western Reserve University, University
of Chicago, Drexel University, Fermilab, the Institute for Advanced Study,
the Japan Participation Group, Johns Hopkins University, the Joint Institute
for Nuclear Astrophysics, the Kavli Institute for Particle Astrophysics and
Cosmology, the Korean Scientist Group, the Chinese Academy of Sciences
(LAMOST), Los Alamos National Laboratory, the Max-Planck-Institute for
Astronomy (MPIA), the Max-Planck-Institute for Astrophysics (MPA), New Mexico
State University, Ohio State University, University of Pittsburgh, University
of Portsmouth, Princeton University, the United States Naval Observatory, and
the University of Washington.

\bsp

\label{lastpage}


\begin{thebibliography}{}
%
%
  \bibitem[\protect\citeauthoryear{Baldwin et al.}{2004}]{Baldwin2004}
    Baldwin J.A., Ferland G.J., Korista K.T., Hamann F., LaCluyz\'e A.,
    2004, ApJ, 615, 610
%
  \bibitem[\protect\citeauthoryear{Bruhweiler \& Verner}{2008}]{Bruhweiler2008}
    Bruhweiler F., Verner E., 2008, ApJ, 675, 83
%
  \bibitem[\protect\citeauthoryear{Clowes \& Campusano}{1991}]{Clowes1991}
    Clowes R.G., Campusano L.E., 1991, MNRAS, 249, 218
%
  \bibitem[\protect\citeauthoryear{Clowes et al.}{2012}]{Clowes2012}
    Clowes R.G., Campusano L.E., Graham M.J., S\"ochting I.K., 2012, MNRAS,
    419, 556
%
  \bibitem[\protect\citeauthoryear{Clowes et al.}{2013}]{Clowes2013}
    Clowes R.G., Harris K.A., Raghunathan S., Campusano L.E., S\"ochting I.K.,
    Graham M.J., 2013, MNRAS, 429, 2910
%
  \bibitem[\protect\citeauthoryear{Collin-Souffrin \& Lasota}{1988}]{Collin-Souffrin1988}
    Collin-Souffrin S., Lasota J.-P., 1988, PASP, 100, 1041
%
  \bibitem[\protect\citeauthoryear{Graham, Clowes \& Campusano}
  {Graham et al.}{1996}]{Graham1996}
    Graham M.J., Clowes R.G., Campusano L.E., 1996, MNRAS, 279, 1349
%
  \bibitem[\protect\citeauthoryear{Harris}{2011}]{Harris2011}
    Harris K.A., PhD thesis, Univ. of Central Lancashire (astro-ph/1201.5746)
%
  \bibitem[\protect\citeauthoryear{Harris et al.}{2012}]{Harris2012}
    Harris K.A., Clowes R.G., Williger G.M., Haberzettl L.G.,
    Campusano L.E., 2012, in Boissier S., de Laverny P., Nardetto N.,
    Samadi R., Valls-Gabaud D., Wozniak H., eds, SF2A-2012: Proceedings
    of the Annual meeting of the French Society of Astronomy and Astrophysics,
    p. 469
%
  \bibitem[\protect\citeauthoryear{Meusinger et al.}{2012}]{Meusinger2012}
    Meusinger H., Schalldach P., Scholz R.-D., in der Au A., Newholm M.,
    de Hoon A., Kaminsky B., 2012, A\&A, 541, A77
%
  \bibitem[\protect\citeauthoryear{Newman et al.}{1998}]{Newman1998}
    Newman P.R., Clowes R.G., Campusano L.E., Graham M.J., 1998, in M\"uller V.,
    Gottl\"ober S., M\"ucket J.P., Wambsganss J., eds, Large Scale Structure:
    Tracks and Traces. World Scientific, Singapore, p. 133
%
  \bibitem[\protect\citeauthoryear{Newman}{1999}]{Newman1999}
    Newman P.R., 1999, PhD thesis, Univ. of Central Lancashire
%
  \bibitem[\protect\citeauthoryear{Penston}{1987}]{Penston1987} Penston M.V.,
    1987, MNRAS, 229, 1P
%
  \bibitem[\protect\citeauthoryear{Richards et al.}{2006}]{Richards2006}
    Richards G.T. et al., 2006, AJ, 131, 2766
%
  \bibitem[\protect\citeauthoryear{Ruff et al.}{2012}]{Ruff2012}
    Ruff A.J., Floyd D.J.E., Webster R.L., Korista K.T., Landt H., 2012,
    ApJ, 754:18
%
  \bibitem[\protect\citeauthoryear{Schneider et al.}{2010}]{Schneider2010}
    Schneider D.P. et al., 2010, AJ, 139, 2360
%
  \bibitem[\protect\citeauthoryear{Shen et al.}{2011}]{Shen2011}
    Shen Y. et al., 2011, ApJS, 194:45 
%
  \bibitem[\protect\citeauthoryear{Sigut \& Pradhan}{1998}]{Sigut1998}
    Sigut T.A.A., Pradhan A.K., 1998, ApJ, 499, L139
%
  \bibitem[\protect\citeauthoryear{Sigut \& Pradhan}{2003}]{Sigut2003}
    Sigut T.A.A., Pradhan A.K., 2003, ApJS, 145, 15
%
  \bibitem[\protect\citeauthoryear{Vanden Berk et al.}{2005}]{Vanden-Berk2005}
    Vanden Berk D.E., et al., 2005, AJ, 129, 2047
%
  \bibitem[\protect\citeauthoryear{Weymann et al.}{1991}]{Weymann1991}
    Weymann R.J., Morris S.L., Foltz C.B., Hewett P.C., 1991, ApJ, 373, 23
%
  \bibitem[\protect\citeauthoryear{Wills, Netzer \& Wills}
  {Wills et al.}{1985}]{Wills1985}
    Wills B.J., Netzer H., Wills D., 1985, ApJ, 288, 94
%
  \bibitem[\protect\citeauthoryear{Zhang}{2011}]{Zhang2011}
    Zhang X.-G., 2011, ApJ, 741, 104
%
\end{thebibliography}
\end{document}